# Propagation of Shear-Horizontal Acoustic Waves in a Quartz Plate with a Fluid Layer for Liquid Sensor Application

Bo Liu[1], Qing Jiang[1] and Jiashi Yang[2] (jyang1@unl.edu)
[1]Department of Mechanical Engineering, University of California, Riverside, CA 92521, USA
[2]Department of Mechanical and Materials Engineering, University of Nebraska, Lincoln, NE 68588-0526, USA

**Abstract**
 We study the propagation of shear-horizontal (SH) waves in a quartz plate in contact with a viscous fluid layer of a finite thickness as an acoustic wave sensor for measuring fluid viscosity or density. Mindlin's first-order plate theory and the theory of Newtonian fluids are used. An equation for determining the dispersion relations of the SH waves is obtained. Approximate dispersion relations for long waves are given analytically. Numerical results showing the effects of the fluid on SH wave characteristics are presented.

## 1. Introduction

Resonant frequencies (or waves speed when a propagating wave is considered in a waveguide) of an elastic body can be affected by many effects like a temperature change or initial stresses, etc. Therefore, frequency shifts in a vibrating elastic body (resonator) can be used to make various so-called acoustic wave sensors. Sometimes piezoelectric crystals are used so that the operating mode or wave can be directly excited and detected electrically. Otherwise transducers are needed for the excitation and detection of acoustic waves.

Specifically, a resonator when in contact with a viscous fluid changes its resonance frequencies due to the inertia and viscosity of the fluid. This effect has been used to make fluid sensors for measuring fluid viscosity or density [1-3]. More references can be found in relevant review articles [4,5]. For fluid sensor applications, vibration modes of an elastic body without a normal displacement at its surface are ideal. In these modes, no pressure waves are generated in the fluid. The fluid produces a drag only on the body surface due to the fluid viscosity and the tangential motion of the surface, thereby lowering the resonance frequencies of the body.

Certain shear modes in plates called thickness-shear (TSh) are widely used for fluid sensor applications [1-3]. Theoretically, these modes can only exist in unbounded plates without edge effects. In these modes, motions of material particles are parallel to the surfaces of the plates, and particle velocities only vary along the plate thickness direction, without in-plane variations. From the viewpoint of wave propagation, TSh modes in a plate are waves propagating along the thickness direction of the plate and are bounced back and forth between the surfaces of the plate. The wave vector is parallel to the plate thickness direction, and the in-plane wave numbers are zero or the in-plane wave lengths are infinite. These TSh waves or modes are the idealized operating modes of many acoustic wave devices.

In reality, however, due to the finite size of devices, pure TSh modes cannot exist because of edge effects. Therefore, in real devices, usually the operating modes are in fact related to waves whose wave vectors have a small in-plane component. These waves have been referred to as essentially TSh waves, or transversely varying TSh waves. In the case when the transverse variation is in a direction perpendicular to the TSh particle velocity, the corresponding waves are called thickness-twist (TT) waves. These transversely varying waves are long waves in plates whose in-plane wave lengths are much larger than the plate thickness. Both stationary waves in resonators and propagating waves in waveguides have been used for resonators and sensors. Understanding the behavior of long waves in plates is fundamentally important to plate acoustic wave devices.

The propagation of waves in crystal plates, elastic or piezoelectric, has been an active research subject for a long time [6-11]. In particular, waves in plates in contact with a fluid have been studied for fluid sensor applications, e.g., [12,13]. Due to material anisotropy, modeling of crystal devices using the three-dimensional (3D) theory of elasticity or piezoelectricity usually involves considerable mathematical difficulties. In fluid sensor applications, this is further complicated by the fluid-structure interaction. Although long equations for determining the wave dispersion relations can often be formulated, they are typically involved with transcendental equations with multi-valued solutions and complex roots. Therefore, numerical search for the roots of the frequency equation is usually needed which presents various challenges.

For the analysis of plate acoustic wave devices, in a series of papers [14-17], Mindlin and his coworkers developed and refined two-dimensional (2D) equations for motions of crystal plates. The 2D plate equations make theoretical and numerical analysis possible in many practically useful cases [18-23]. The 2D equations effectively reduce the dimension of the problem by one, which is a significant simplification. In addition, the plate equations usually are only involved with the particular operating mode of a device plus a few other modes that are coupled to the mode of interest. Therefore, mode identification when using 2D equations is much simpler than when using 3D equations.

In this paper, we use Mindlin's first-order plate equations to study SH waves in a quartz crystal plate in contact with a viscous fluid layer. In addition to pure TSh modes [24,25], we are mainly interested in the propagation of long waves and how they are affected by the presence of the fluid [26]. The fluid in [26] occupies a semi-infinite space. In this paper, we study the effect of a fluid layer with a finite thickness. The use of 2D plate equations simplifies the problem and allows us to obtain some simple, fundamental and analytical results useful to the understanding and design of plate wave fluid sensors.

## 2. Two-Dimensional Plate Equations

The equations for crystal plates vary considerably according to the symmetry of the crystals. Quartz is a crystal widely used for acoustic wave devices. Therefore we focus on quartz in the following. Quartz has very weak piezoelectric coupling. For frequency analysis, the small piezoelectric coupling may be neglected and an elastic analysis is usually sufficient. This is common practice in the frequency analysis of resonant quartz devices. A particular cut of a quartz plate refers to the orientation of the plate when it is taken out of an anisotropic bulk quartz crystal. As a consequence, quartz plates of different cuts exhibit different anisotropies in coordinates normal and parallel to the plate surfaces. Rotated Y-cut quartz plates are effectively monoclinic. They include the most frequently used AT-cut quartz plates as a special case. In this section, we summarize the 2D plate equations for rotated Y-cut quartz [14,16]. Consider such a plate as shown in Fig. 1. It is in contact with a Newtonian fluid layer of thickness $H$. For rotated Y-cut quartz plates, SH (or antiplane) motions with only one displacement component are allowed by the linear theory of anisotropic elasticity. These motions are particularly useful in device applications. They are described by

$$u_1 = u_1(x_2, x_3, t), \quad u_2 = u_3 = 0, \qquad (1)$$

where **u** is the displacement vector. $u_1$ is governed by [27]

$$c_{66} u_{1,22} + c_{55} u_{1,33} + 2 c_{56} u_{1,23} = \rho \ddot{u}_1. \qquad (2)$$

Exact solutions to (2) may be attempted for relatively simple problems. Due to $c_{56}$, solving (2) is not straightforward and the results are usually complicated. (2) includes all SH modes. Since an acoustic wave device usually operates with a particular mode, it is simper to use 2D plate equations which describe the modes of interest only. The ideal operating mode is TSh which does not have $x_1$ and $x_3$ dependence. The variation of these modes along $x_1$ has been reasonably well understood. Therefore this paper is concerned with the $x_3$ dependence only without the $x_1$ dependence, i.e., the so-called straight-crested waves or modes. For straight-crested SH motions, the displacement field of the first-order plate theory is approximated by [14,16]

$$u_1(x_2, x_3, t) = u_1^{(0)}(x_3, t) + x_2 u_1^{(1)}(x_3, t), \qquad (3)$$

where $u_1^{(0)}(x_3, t)$ is the face-shear (FS) displacement, and $u_1^{(1)}(x_3, t)$ is the fundamental TT displacement. When $u_1^{(1)}(x_3, t)$ is independent of $x_3$, it reduces to the fundamental TSh displacement which is the ideal mode of an infinite plate. Both $u_1^{(0)}$ and $u_1^{(1)}$ have tangential surface displacements only and are very useful for fluid sensor application. They are governed by the following plate equations of motion [14,16]:

$$T_{31,3}^{(0)} + 2bT_1^{(0)} = 2b\rho \ddot{u}_1^{(0)}, \qquad (4a)$$

$$T_{31,3}^{(1)} - T_{21}^{(0)} + \frac{2b^3}{3} T_1^{(1)} = \frac{2b^3}{3} \rho \ddot{u}_1^{(1)}. \qquad (4b)$$

The plate resultants $T_{31}^{(0)}$, $T_{21}^{(0)}$ and $T_{31}^{(1)}$ over a cross section of the plate represent plate internal forces and moments. They are related to the plate displacements $u_1^{(0)}$ and $u_1^{(1)}$ by the following constitutive relations [14,16]:

$$T_{31}^{(0)} = 2b(c_{55} u_{1,3}^{(0)} + \kappa_1 c_{56} u_1^{(1)}), \qquad (5a)$$

$$T_{12}^{(0)} = 2b(\kappa_1 c_{56} u_{1,3}^{(0)} + \kappa_1^2 c_{66} u_1^{(1)}), \qquad (5b)$$

$$T_{31}^{(1)} = \frac{2b^3}{3}(\gamma_{55} u_{1,3}^{(1)} + \psi_{35} \phi_{,3}^{(1)}), \qquad (5c)$$

where $c_{pq}$ is the usual elastic stiffness, $\gamma_{55} = 1/s_{55}$, and $s_{pq}$ is the elastic compliance. $\kappa_1$ in the above equations is a shear correction factor [14,16] which will be determined later. The mechanical surface loads in (4) are defined by

$$T_1^{(0)} = \frac{1}{2b}\left[T_{21}(b^-) - T_{21}(-b^-)\right],$$

$$T_1^{(1)} = \frac{3}{2b^3}\left[bT_{21}(b^-) + bT_{21}(-b^-)\right], \qquad (6)$$

where $b^-$ is the lower limit of $b$. The substitution of (5) into (4) gives two equations for $u_1^{(0)}$ and $u_1^{(1)}$:

$$c_{55} u_{1,33}^{(0)} + \kappa_1 c_{56} u_{1,3}^{(1)} + T_1^{(0)} = \rho \ddot{u}_1^{(0)}, \qquad (7a)$$

$$\gamma_{55} u_{1,33}^{(1)} - 3b^{-2}(\kappa_1 c_{56} u_{1,3}^{(0)} + \kappa_1^2 c_{66} u_1^{(1)}) + T_1^{(1)} = \rho \ddot{u}_1^{(1)}. \qquad (7b)$$

Clearly, $c_{56}$ causes the coupling between $u_1^{(0)}$ and $u_1^{(1)}$. Therefore, a coupled analysis is necessary. (7) has spatial derivatives with respective to $x_3$ only but not $x_2$ due to the plate approximation, and therefore is much simpler than (2).

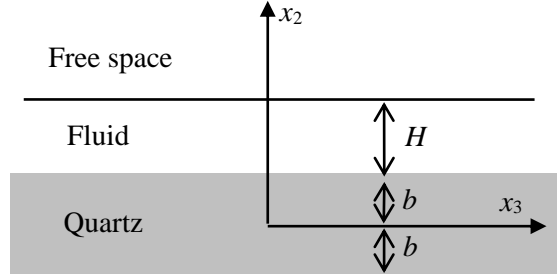

Free space

Fig. 1. A crystal plate with a fluid layer

### 3. Thickness-shear Vibration

The shear correction factor $\kappa_1$ in the plate equations in the previous section is determined by requiring the resonance frequency of the fundamental TSh mode calculated from the 3D exact equations and the 2D plate equations to be the same [14,16]. The relevant 3D solution was given in [24]. In this section we calculate the 2D solution and determine $\kappa_1$. Consider an unbounded plate. Whether the fluid is compressible or not does not matter because the motion to be considered is a pure shear without volume change. For time-harmonic motions, we use the usual complex notation. All fields have the same $\exp(i\omega t)$ factor which will be dropped below for simplicity.

**3.1 Fluid**

The equation of motion for the fluid is [28]

$$T_{21,2} = \rho_L \dot{v}_1, \tag{8}$$

where the shear stress is given by

$$T_{21} = \mu \frac{\partial v_1}{\partial x_2} . \tag{9}$$

$\mu$ and $\rho_L$ are the viscosity and mass density of the fluid. $v_1$ and $T_{21}$ are the relevant velocity and shear stress components. The velocity field can be determined as

$$v_1 = \{C_1 \sinh[(1+i)\eta(x_2 - b)] + C_2 \cosh[(1+i)\eta(x_2 - b)]\}, \tag{10}$$

where $C_1$ and $C_2$ are undetermined constants, and

$$\eta = \sqrt{\frac{\rho_L \omega}{2\mu}} . \tag{11}$$

The shear stress needed for boundary and continuity conditions is

$$T_{21} = (1+i)\mu\eta\{C_1 \cosh[(1+i)\eta(x_2 - b)] + C_2 \sinh[(1+i)\eta(x_2 - b)]\}. \tag{12}$$

**3.2 Crystal Plate**

For thickness vibrations independent of $x_3$, with the use of (6), (7) reduces to

$$\frac{1}{2b}\left[T_{21}(b^-) - T_{21}(-b^-)\right] = \rho \ddot{u}_1^{(0)},$$

$$-3b^{-2}\kappa_1^2 c_{66} u_1^{(1)} + \frac{3}{2b^3}\left[bT_{21}(b^-) + bT_{21}(-b^-)\right] = \rho \ddot{u}_1^{(1)}. \tag{13}$$

The bottom of the plate surface is traction free, with $T_{21}(-b^-)=0$. At the top of the plate the shear stress is continuous, i.e., $T_{21}(b^-)=T_{21}(b^+)$ where $b^+$ is the upper limit of $b$. With these, (13) becomes

$$\frac{1}{2b}T_{21}(b^+) = \rho \ddot{u}_1^{(0)},$$
$$-3b^{-2}\kappa_1^2 c_{66} u_1^{(1)} + \frac{3}{2b^2}T_{21}(b^+) = \rho \ddot{u}_1^{(1)}. \tag{14}$$

We let

$$u_1^{(0)} = C_3 \exp(i\omega t), \quad u_1^{(1)} = C_4 \exp(i\omega t) \tag{15}$$

where $C_3$ and $C_4$ are undetermined constants.

### 3.3 Boundary and Continuity Conditions
At the top of the fluid layer, we have the following traction-free condition:
$$T_{21}(b+H) = 0. \tag{16}$$
At the interface between the crystal plate and the fluid, we have the continuity of particle velocity:
$$\dot{u}_1^{(0)} + b\dot{u}_1^{(1)} = v_1(b^+). \tag{17}$$
The substitution of the relevant fields in (10), (12), and (15) into (14), (16) and (17) results in four linear and homogeneous equations for $C_1$ through $C_4$:

$$\cosh[(1+i)\eta H]C_1 + \sinh[(1+i)\eta H]C_2 = 0,$$
$$C_2 - i\omega C_3 - i\omega b C_4 = 0,$$
$$(1+i)\mu\eta C_1 + 2\rho b\omega^2 C_3 = 0, \tag{18}$$
$$3(1+i)\mu\eta C_1 + (2\rho b^2 \omega^2 - 6\kappa_1^2 c_{66})C_4 = 0.$$

For nontrivial solutions the determinant of the coefficient matrix of (18) has to vanish, which gives the following frequency equation:

$$\omega^2 - \frac{3\kappa_1^2 c_{66}}{\rho b^2} = \frac{(1-i)\mu\eta}{2\rho b\omega}\left(\frac{3\kappa_1^2 c_{66}}{\rho b^2} - 4\omega^2\right)\tanh[(1+i)\eta H]. \tag{19}$$

### 3.4 Correction Factor
The exact fundamental TSh frequency from the 3D equations when the fluid is not present is given by [29]

$$\omega_0 = \frac{\pi}{2b}\sqrt{\frac{c_{66}}{\rho}}. \tag{20}$$

When the plate is in contact with a low viscosity fluid layer, the fundamental TSh frequency is approximately given by [24]

$$\omega = \omega_0(1+\Delta\Omega), \tag{21}$$

where [24]

$$\Delta\Omega = -\frac{1-i}{\pi}\sqrt{\frac{\rho_L \mu \omega_0}{2\rho c_{66}}}\tanh[(1+i)\eta_0 H], \tag{22}$$

$$\eta_0 = \sqrt{\rho_L \omega_0/(2\mu)}. \tag{23}$$

Substitution of (21) into (19) determines

$$\kappa_1^2 = \frac{\pi^2}{12}(1-\Delta\Omega). \qquad (24)$$

We note that (24) is complex. Its real part is a frequency shift. Its imaginary part describes damped modes due to viscosity.

## 4. Propagation of Face-shear and Thickness-twist Waves

With $\kappa_1$ determined, the plate equations are ready to be used to study propagating waves in the plate which is the main purpose of this paper. We begin with coupled FS and TT waves and then examine uncoupled long FS and long TT waves separately.

### 4.1 Coupled Waves

For propagating waves with both $x_2$ and $x_3$ dependence, the equations for the fluid are
$$T_{21,2} + T_{31,3} = \rho_L \dot{v}_1,$$
$$T_{21} = \mu \frac{\partial v_1}{\partial x_2}, \quad T_{31} = \mu \frac{\partial v_1}{\partial x_3}. \qquad (25)$$

Substituting the stresses in (25) into the equation of motion in (25) gives
$$\mu(v_{1,22} + v_{1,33}) = \rho_L \dot{v}_1. \qquad (26)$$

We consider the following propagating waves:
$$v_1 = \left\{ C_1 \sinh\left[\sqrt{\zeta^2 - 2i\eta^2}(x_2-b)\right] + C_2 \cosh\left[\sqrt{\zeta^2 - 2i\eta^2}(x_2-b)\right]\right\} \exp[i(\zeta x_3 - \omega t)], \qquad (27)$$

where $C_1$ and $C_2$ are undetermined constants, and (11) is still valid. For propagating waves in the crystal plate, we have
$$c_{55}u_{1,33}^{(0)} + \kappa_1 c_{56}u_{1,3}^{(1)} + \frac{1}{2b}T_{21}(b^+) = \rho \ddot{u}_1^{(0)},$$
$$\gamma_{55}u_{1,33}^{(1)} - 3b^{-2}(\kappa_1 c_{56}u_{1,3}^{(0)} + \kappa_1^2 c_{66}u_1^{(1)}) + \frac{3}{2b^3}T_{21}(b^+) = \rho \ddot{u}_1^{(1)}. \qquad (28)$$

At the top of the fluid layer the traction-free boundary condition in (16) still holds. The continuity of velocity at the top of the plate surface given by (17) is still valid too. Let
$$u_1^{(0)} = C_3 \exp[i(\zeta x_3 - \omega t)], \quad u_1^{(1)} = C_4 \exp[i(\zeta x_3 - \omega t)], \qquad (29)$$

where $C_3$ and $C_4$ are undetermined constants. The substitution of (27) and (29) into (16), (17) and (28) yields the following four linear and homogeneous equations for $C_1$ through $C_4$:
$$C_1 + C_2 \tanh\left(\sqrt{\zeta^2 - 2i\eta^2}H\right) = 0,$$
$$C_2 + i\omega C_3 + ib\omega C_4 = 0,$$
$$\frac{\mu\sqrt{\zeta^2 - 2i\eta^2}}{2b}C_1 + (\rho\omega^2 - c_{55}\zeta^2)C_3 + i\kappa_1 c_{56}\zeta C_4 = 0, \qquad (30)$$
$$\frac{3}{2}\mu\sqrt{\zeta^2 - 2i\eta^2}C_1 - 3i\kappa_1 c_{56}\zeta C_3 + (\rho b^2 \omega^2 - \gamma_{55}b^2\zeta^2 - 3\kappa_1^2 c_{66})C_4 = 0.$$

For nontrivial solutions, the determinant of the coefficient matrix has to vanish, which, together with (11), gives the flowing frequency equation that determines the dispersion relation of $\omega$ versus $\zeta$:
$$(iAB\omega + \rho\omega^2 - c_{55}\zeta^2)(3iABb^2\omega + \rho b^2\omega^2 - \gamma_{55}b^2\zeta^2 - 3\kappa_1^2 c_{66})$$
$$- (iABb\omega + i\kappa_1 c_{56}\zeta)(3iABb\omega - 3i\kappa_1 c_{56}\zeta) = 0, \qquad (31)$$

where

$$A = \frac{\mu\sqrt{\zeta^2 - 2i\eta^2}}{2b}, \quad B = \tanh\left(\sqrt{\zeta^2 - 2i\eta^2}\,H\right). \tag{32}$$

In device applications usually long waves with a small or infinitesimal $\zeta$ are used. Therefore, we expand the relevant terms in (31) into power series of $\zeta$ and neglect powers higher than two. Then (31) may be written as

$$F(\omega)\zeta^2 + G(\omega) = 0, \tag{33}$$

where

$$\begin{aligned}F(\omega) &= 4A_1B_2i\rho b^2\omega^3 + 4A_2B_1i\rho b^2\omega^{5/2} - A_1B_1i\gamma_{55}b^2\omega^{3/2} \\ &\quad - 3A_1B_2i\kappa_1^2c_{66}\omega - 3A_2B_1i\kappa_1^2c_{66}\omega^{1/2} - 3A_1B_1ic_{55}b^2\omega^{3/2} \\ &\quad - \gamma_{55}\rho b^2\omega^2 - c_{55}\rho b^2\omega^2 + 3\kappa_1^2c_{66}c_{55} - 3\kappa_1^2c_{56}^2, \\ G(\omega) &= 4A_1B_1i\rho b^2\omega^{7/2} - 3A_1B_1i\kappa_1^2c_{66}\omega^{3/2} + \rho^2b^2\omega^4 - 3\kappa_1^2c_{66}\rho\omega^2,\end{aligned} \tag{34}$$

and

$$A_1 = \frac{1}{2b}\sqrt{-i\mu\rho_L}, \quad A_2 = \frac{1}{4b}\sqrt{\frac{i\mu^3}{\rho_L}},$$

$$B_1(\omega) = \tanh\left(\sqrt{-2i\eta^2}\,H\right), \quad B_2(\omega) = \frac{H\left[1 - \tanh^2\left(\sqrt{-2i\eta^2}\,H\right)\right]}{2}\sqrt{\frac{i\mu}{\rho_L}}. \tag{35}$$

In the special case when the fluid is not present, for thickness modes with $\zeta=0$, (33) determines two frequencies of 0 and $\omega_0$. For small but nonzero values of $\zeta$, (33) determines two dispersion relations for FS and TT waves. The FS branch goes through the origin $(\omega,\zeta)=(0,0)$. The TT branch has a finite intercept at $(\omega,\zeta)=(\omega_0,0)$ and $\omega_0$ is called the cutoff frequency below which the TT wave becomes a exponential function of $x_3$ and cannot propagate. When the fluid is present, for low-viscosity fluids, we expect the two branches are modified slightly and discuss them separately below.

### 4.2 Long FS Waves

For the FS branch, when $\zeta$ is small, $\omega$ is also small and is of the same order (see [30] for the case when the fluid is not present). In this case, neglecting higher powers of $\omega$, (33) may be approximated by

$$(3\kappa_1^2c_{66}c_{55} - 3\kappa_1^2c_{56}^2)\zeta^2 = 3\kappa_1^2c_{66}\rho\omega^2 + 3A_1B_1i\kappa_1^2c_{66}\omega^{3/2}, \tag{36}$$

where the effect of the fluid is represented by the second term on the right-hand side. When the fluid is not present and the crystal plate is alone, denoting the wave frequency by $\omega_{\text{Plate}}$, (36) reduces to the following known dispersion relation for long FS waves [30]:

$$\omega_{\text{Plate}} = \sqrt{\frac{c_{66}c_{55} - c_{56}^2}{\rho c_{66}}}\,\zeta = \sqrt{\frac{\gamma_{55}}{\rho}}\,\zeta, \tag{37}$$

which is nondispersive. An immediate observation is that the second term on the right-hand side of (36) makes the waves dispersive. For numerical results, an AT-cut quartz plate with $b=1$mm is used, which is a typical thickness for quartz devices. For the fluid, we choose as an example chloroform with $\rho_L = 1.483\times 10^3\,\text{kg/m}^3$ and a relatively low viscosity of $\mu = 0.542\,\text{mPa}\cdot\text{s}$ (smaller than the viscosity of water). We consider real frequencies and solve (36) for complex wave numbers. For figure plotting we introduce the following dimensionless frequencies and wave number:

$$\Omega = \frac{\omega}{\omega_0}, \quad \Omega_{\text{Plate}} = \frac{\omega_{\text{Plate}}}{\omega_0}, \quad Z = \frac{\zeta}{\pi/(2b)} = X + iY. \tag{38}$$

Long waves are described by small values of the real part of $Z$.

Figure 2 shows the effect of viscosity on the dispersion relation of long FS waves. The viscosity of chloroform is artificially varied while other parameters are fixed. Figure 2 (a) shows a fundamental and qualitative effect of the fluid viscosity which changes the dispersion curve from real to complex indicating attenuation due to fluid viscosity, and from a straight line to a curve showing dispersion. Figure 2 (b) shows the real part of the dispersion relation which is essentially linear and hence the fluid induced dispersion is small. Figure 2 (c) shows the difference between the real parts of (36) and (37), in which the fluid induced dispersion becomes visible. It can be seen from Fig. 2 (c) that, although long waves are usually used in real devices, the frequency shift is larger when the wave number $X$ is larger or the wave is shorter. This is because shorter waves have lager velocity gradients in the $x_3$ direction and hence a larger viscous stress $T_{31}$. Figure 2 (c) also shows that the fluid lowers the frequency, and higher viscosity causes more frequency shift. The relative frequency shift is of the order of $10^{-5}$ which is considered a significant frequency change because the thermal noise in crystal resonators is typically of the order of $10^{-6}$. Therefore a $10^{-5}$ frequency shift is a clear and measurable signal. Figure 2 (d) shows the imaginary part of the complex wave number. It is positive, representing attenuation for the right-traveling waves given in (29). For a fixed frequency, higher viscosity causes larger attenuation.

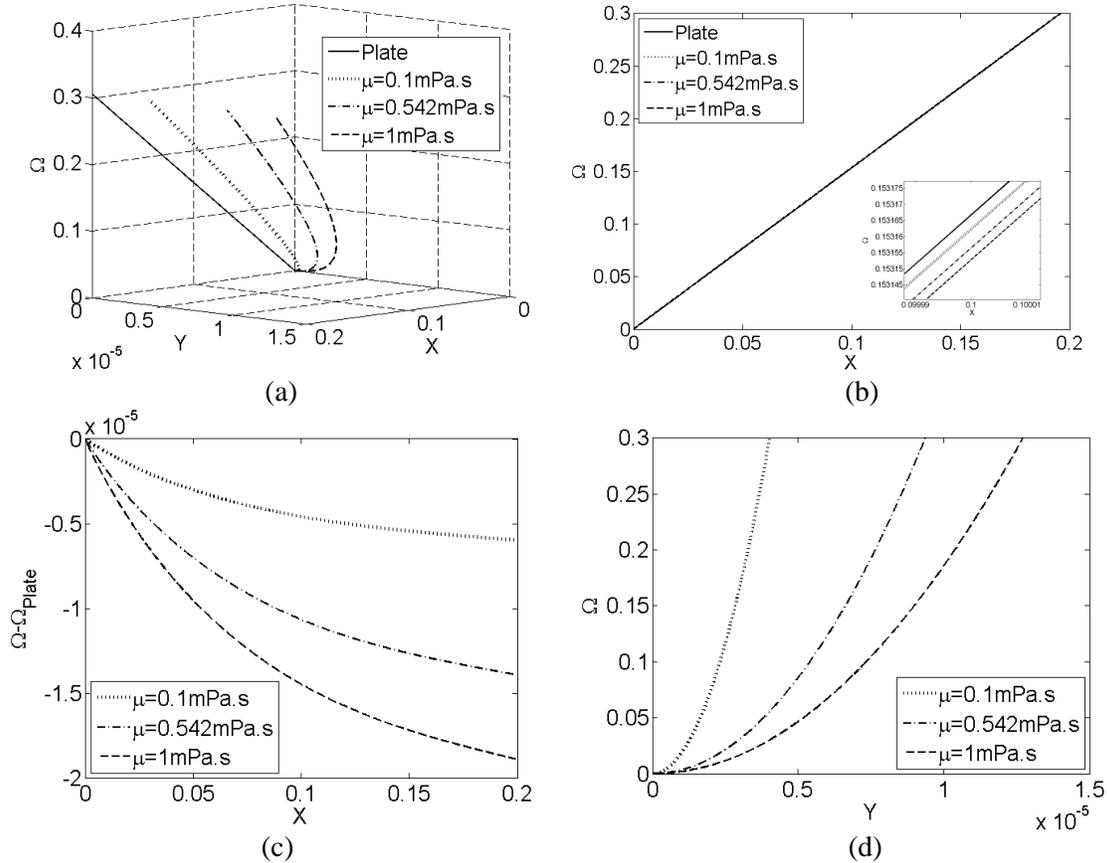

Fig. 2. Effects of fluid viscosity on FS waves, $H=2b$

Figure 3 shows the effect of the fluid density. The behaviors in Fig. 3 are similar to those in Fig. 2. This is as expected because, as shown in (22), the fluid viscosity and density appear together in a product in the first-order approximation of the frequency shift. In a typical application, one needs to know either the fluid density or viscosity and then uses an acoustic wave fluid sensor to measure the other. How to separate the density from viscosity is a challenging problem in acoustic wave fluid sensors.

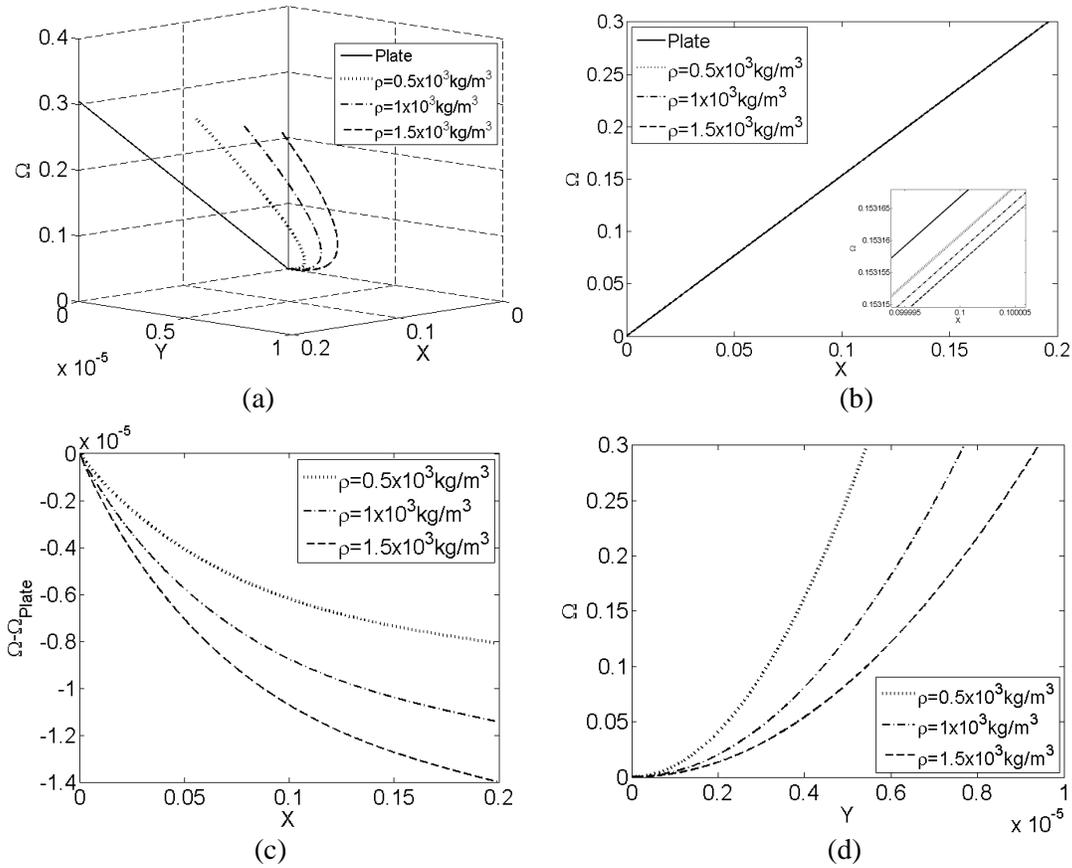

Fig. 3. Effects of fluid density on FS waves, $H=2b$

Figure 4 shows the effect of the fluid layer thickness $H$. The cases of $H=0.5b$ and $H=1b$ are indistinguishable in the figure. In this case, effectively, the fluid layer can be treated as a semi-infinite half space [26]. The frequency shift is an increasing function of $H$ for small H only [24]. When $H$ reaches a certain value, there exists a maximum frequency shift after which the frequency shift decreases with $H$ [24]. What is shown in Fig. 4 (c) is the case when a large $H$ has a small frequency shift.

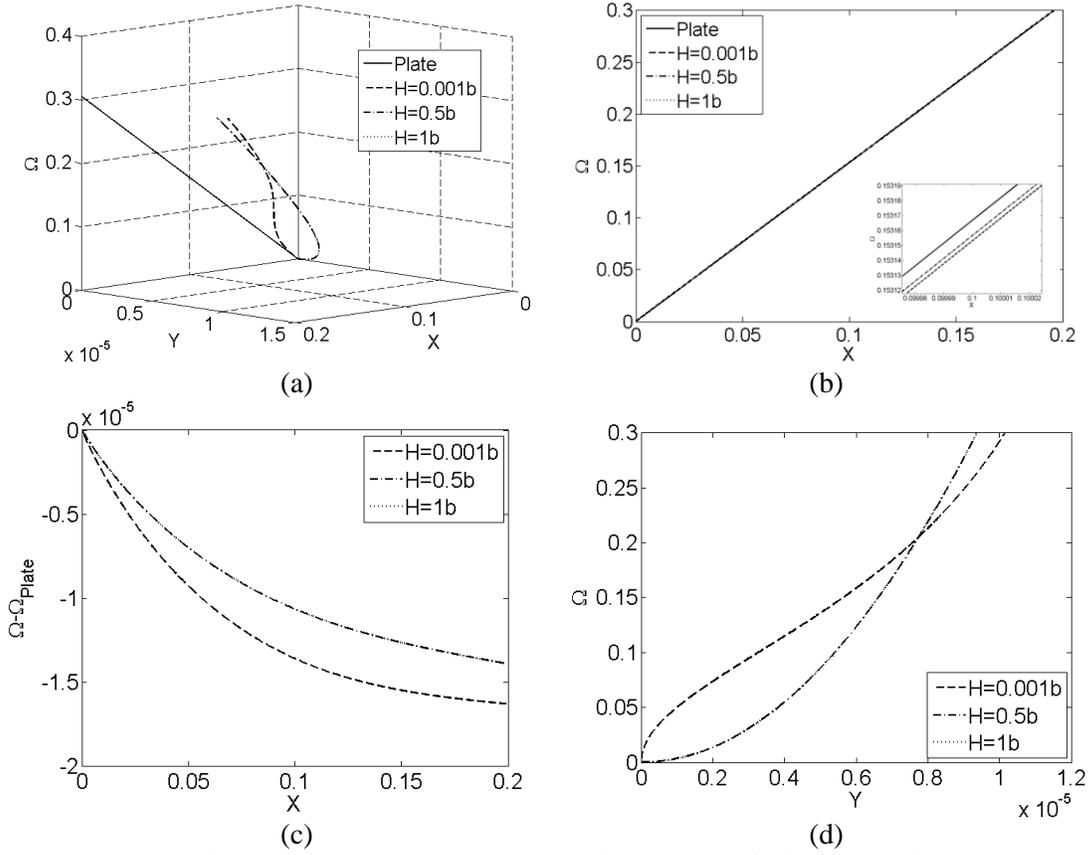

Fig. 4. Effects of fluid layer thickness on FS waves ($H=0.5b$ and $H=1b$ overlap)

**4.3 Long TT Waves**

For the TT branch of the dispersion relations, when $\zeta$ is small, $\omega$ is finite. In this case, we denote

$$\omega = \omega_0(1+\Delta\Omega), \tag{39}$$

where $\Delta\Omega$ is small. Substituting (39) into (31), for small $\Delta\Omega$, we obtain

$$\frac{\omega-\omega_0}{\omega_0} = \Delta\Omega = -\frac{D_3}{D_4} - \frac{D_1 D_4 - D_2 D_3}{D_4^2}\zeta^2, \tag{40}$$

where

$$\begin{aligned}
D_1 &= 4A_1 B_2^0 i\rho b^2 \omega_0^3 + 4A_2 B_1^0 i\rho b^2 \omega_0^{5/2} - A_1 B_1^0 i\gamma_{55} b^2 \omega_0^{3/2} \\
&\quad - 3A_1 B_2^0 i\kappa_1^2 c_{66}\omega_0 - 3A_2 B_1^0 i\kappa_1^2 c_{66}\omega_0^{1/2} - \gamma_{55}\rho b^2 \omega_0^2 - 3A_1 B_1^0 i c_{55} b^2 \omega_0^{3/2} \\
&\quad - c_{55}\rho b^2 \omega_0^2 + 3\kappa_1^2 c_{66} c_{55} - 3\kappa_1^2 c_{56}^2, \\
D_2 &= 12 A_1 B_2^0 i\rho b^2 \omega_0^3 + 10 A_2 B_1^0 i\rho b^2 \omega_0^{5/2} - \frac{3}{2} A_1 B_1^0 i\gamma_{55} b^2 \omega_0^{3/2} - 3A_1 B_2^0 i\kappa_1^2 c_{66}\omega_0 \\
&\quad - \frac{3}{2} A_2 B_1^0 i\kappa_1^2 c_{66}\omega_0^{1/2} - 2\gamma_{55}\rho b^2 \omega_0^2 - \frac{9}{2} A_1 B_1^0 i c_{55} b^2 \omega_0^{3/2} - 2c_{55}\rho b^2 \omega_0^2, \\
D_3 &= 4A_1 B_1^0 i\rho b^2 \omega_0^{7/2} - 3A_1 B_1^0 i\kappa_1^2 c_{66}\omega_0^{3/2} + \rho^2 b^2 \omega_0^4 - 3\kappa_1^2 c_{66}\rho\omega_0^2, \\
D_4 &= 14 A_1 B_1^0 i\rho b^2 \omega_0^{7/2} - \frac{9}{2} A_1 B_1^0 i\kappa_1^2 c_{66}\omega_0^{3/2} + 4\rho^2 b^2 \omega_0^4 - 6\kappa_1^2 c_{66}\rho\omega_0^2.
\end{aligned} \tag{41}$$

In (41), for low-viscosity fluids, since $B_1$ and $B_2$ are always multiplied with $A_1$ or $A_2$ which depend on $\mu$, $B_1$ and $B_2$ have been approximated by

$$B_1^0(\omega) \cong B_1(\omega_0), \quad B_2^0(\omega) \cong B_2(\omega_0). \tag{42}$$

(40) shows that locally, near cutoff, the dispersion curve may be approximated by a parabola and therefore long TT waves are dispersive. When $\zeta = 0$, (40) reduces to

$$\omega = \omega_0 \left(1 - \frac{D_3}{D_4}\right), \tag{43}$$

which is complex. Numerical results for the effects of fluid viscosity, density, and layer thickness on long TT waves are shown in Figs. 5, 6, and 7, respectively. When there is no fluid, part of the dispersion curve of TT waves is real and the rest is purely imaginary. The dispersion curve does not go through the origin. It has a finite intercept with the $\Omega$ axis which is the cutoff frequency. The basic effects of the fluid on long TT waves are similar to the case of FS waves. The dispersion curve becomes complex. TT waves are with higher frequencies than FS waves and therefore decay faster in the fluid.

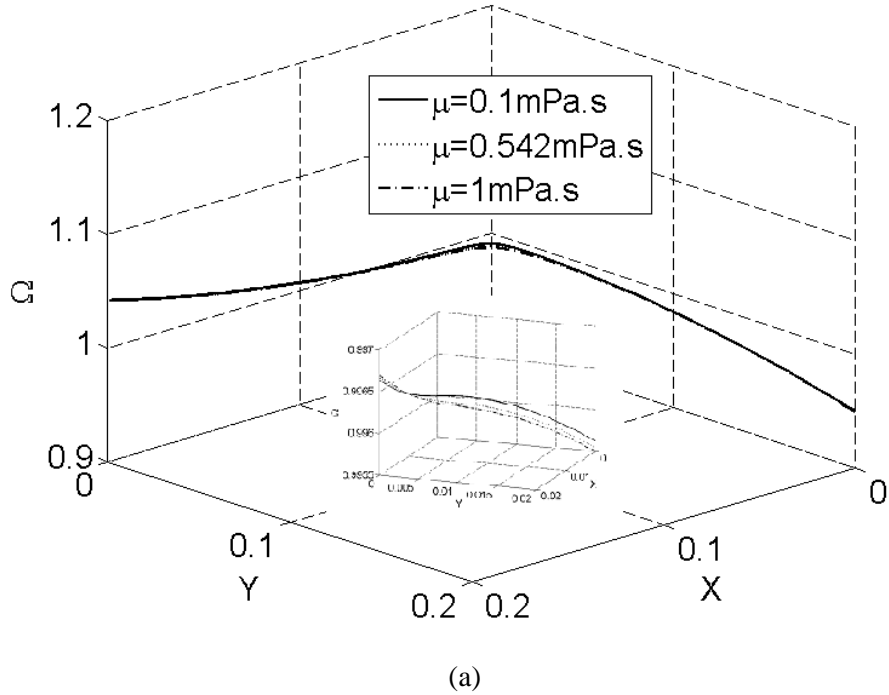

(a)

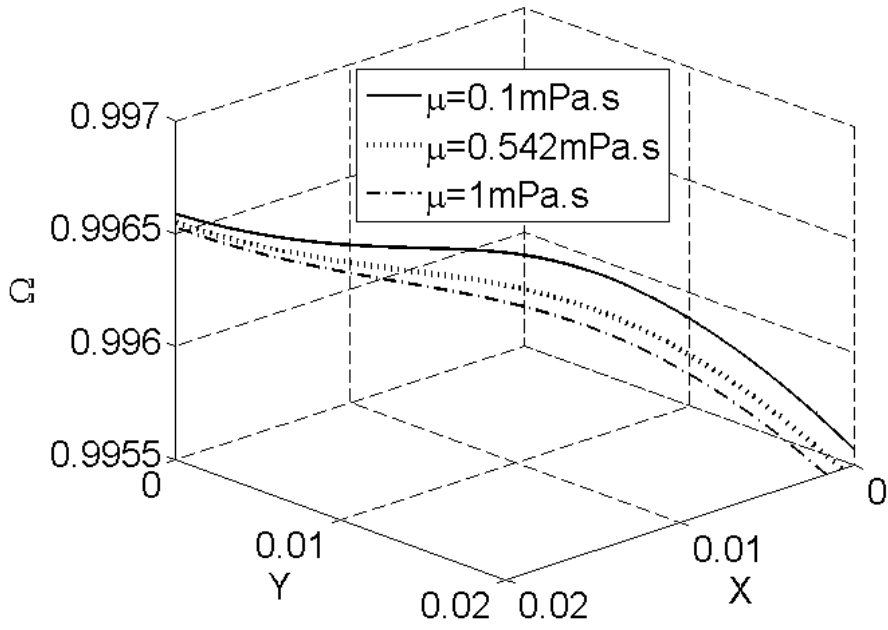

(b)

Fig. 5. Effects of fluid viscosity on TT waves, $H=2b$

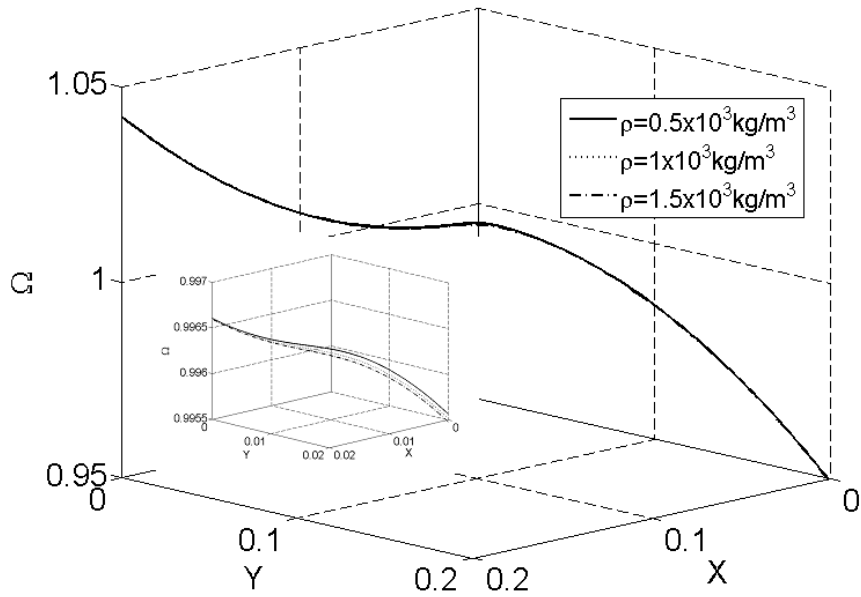

(a)

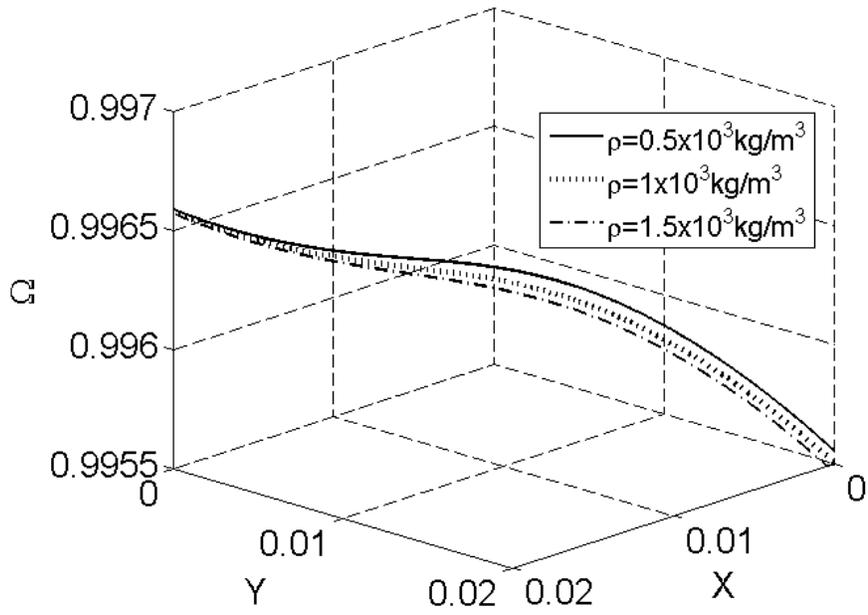

(b)

Fig. 6. Effects of fluid density on TT waves, $H=2b$

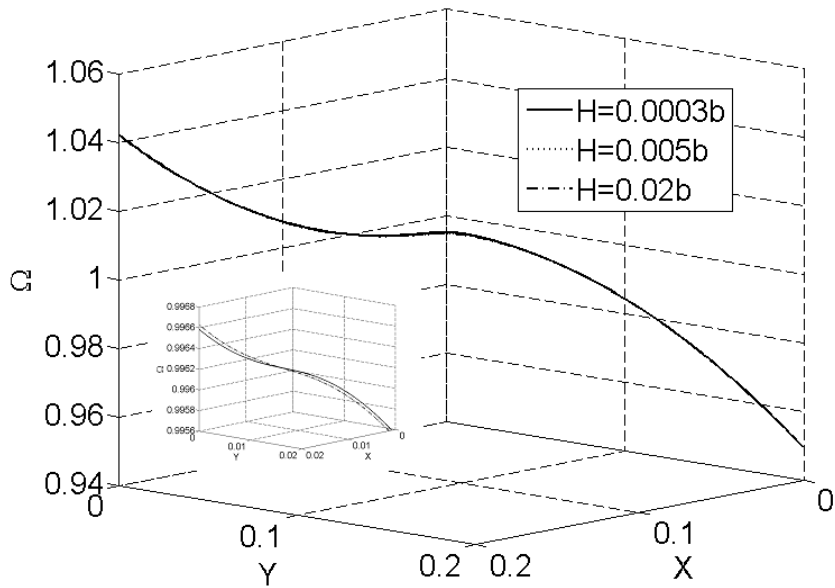

(a)

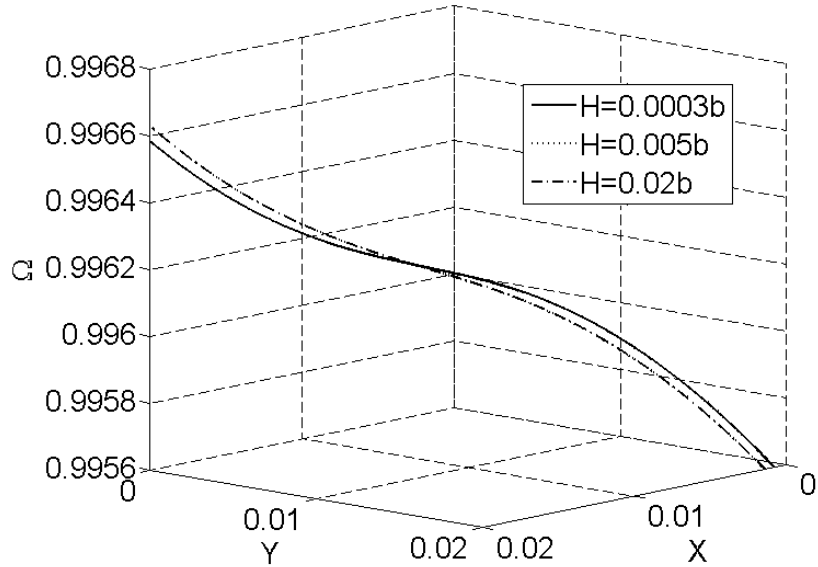

(b)
Fig. 7. Effects of fluid layer thickness on TT waves ($H$=0.005$b$ and $H$=0.02$b$ overlap)

**5. Conclusions**

Analytical solutions are obtained for TSh modes and FS as well as TT waves in a crystal plate carrying a viscous fluid layer with a finite thickness. Approximate expressions for frequency shifts and dispersion relations are presented. The frequencies and dispersion relations become complex due to the fluid, indicating damped modes and waves with attenuation. The fluid viscosity and density lower the frequencies together in a combined manner, causing (additional) dispersion. Shorter waves have larger frequency shifts. Long FS waves become dispersive due to the fluid. Typical relative frequency shifts are of the order of $10^{-5}$ which are detectable by crystal resonators and waveguides. The results obtained are fundamental and useful for the understanding and design of quartz crystal fluid sensors.